\documentclass[prd,aps,twocolumn,showpacs,preprintnumbers,amsmath,amssymb]{revtex4}
\usepackage{mathrsfs}
\usepackage{amsmath,amssymb,graphicx}
\usepackage {amssymb}
\usepackage{color}
\newcommand{\nc}{\newcommand}
\nc{\bea}{\begin{eqnarray}} \nc{\eea}{\end{eqnarray}}
\nc{\be}{\begin{equation}} \nc{\ee}{\end{equation}}

\input{epsf.sty}

\begin{document}

\title{Perturbations of Single-field Inflation in Modified Gravity Theory}

\author{Taotao Qiu}
\email{qiutt@mail.ccnu.edu.cn}
\affiliation{Key Laboratory of Quark and Lepton Physics (MOE) and College of Physical Science $\&$ Technology, Central China Normal University, Wuhan 430079, P.R.China}

\author{Jun-Qing Xia}
\email{xiajq@ihep.ac.cn}
\affiliation{Key Laboratory of Particle Astrophysics, Institute of High Energy Physics, Chinese Academy of Science, P. O. Box 918-3, Beijing 100049, P. R. China}


\begin{abstract}
In this paper, we study the case of single field inflation within the framework of modified gravity theory where the gravity part has an arbitrary form $f(R)$. Via a conformal transformation, this case can be transformed into its Einstein frame where it looks like a two-field inflation model. However, due to the existence of the isocurvature modes in such a multi-degree-of-freedom (m.d.o.f.) system, the (curvature) perturbations are not equivalent in two frames, so in despite of its convenience, it is illegal to treat the perturbations in its Einstein frame as the ``real" ones as we always do for pure $f(R)$ theory or single field with nonminimal coupling. Here by pulling the results of curvature perturbations back into its original Jordan frame, we show explicitly the power spectrum and spectral index of the perturbations in the Jordan frame, as well as how it differs from the Einstein frame. We also fit our results with the newest Planck data. Since there are large parameter space in these models, we show that it is easy to fit the data very well.
\end{abstract}
\maketitle

\section{Introduction}

As a cosmological paradigm of the early universe, inflation has achieved many reputations thanks to its capability in addressing several problems of Big Bang cosmology \cite{Guth:1980zm, Linde:1981mu, Albrecht:1982wi} such as horizon, flatness, unwanted relics and so on (see \cite{Starobinsky:1980te, Fang:1980wi, Sato:1980yn} for early works). Furthermore, inflation predicts a nearly scale-invariant primordial power spectrum, which was verified to high precision by Cosmic Microwave Background (CMB) observations \cite{Komatsu:2010fb, Ade:2013lta}. Due to these salient features, more and more people has been paying attention to the study of inflation and more and more models has been explored. Observational data has also been developed to better and better precision in order to verify/falsify these models.

Among a variety of theoretical models of inflation, one interesting kind of models is that based on modified gravity. In these models, Einstein's Gravity is modified in high energy level, which can lead to a period of inflationary expansion of the universe. One of the example is the well-known Starobinsky inflation model \cite{Starobinsky:1980te}, where a squared term of Ricci scalar has been added to the normal Lagrangian of Gravity. The newest PLANCK constraints on inflation has shown us that the Starobinsky model is among the best fit models to the observations \cite{Ade:2013lta,Planck:2015xua}. However, there may also exist some scalar fields at early times of the universe (such as axions \cite{Peccei:1977hh} or curvaton field \cite{Lyth:2001nq}), and observations do not exclude the possibilities that the observed CMBR fluctuations come from more than one degree of freedom \cite{Ade:2013lta,Planck:2015xua}. Therefore as a generic study, we want to extend the modified gravity inflation models by adding an additional scalar field. Note that Ref. \cite{Domenech:2015qoa} comes out during the revision of this paper, which studies similar model while taking the scalar field as a curvaton (see also \cite{Cai:2013caa} for earlier works). Quantum corrections of this type of models have also been studied in e.g. \cite{Bamba:2014jia}.


In usual modified gravity models, it is convinient to transform them from its original Jordan frame to the Einstein frame, where they look like field theory models minimally coupling to gravity, which is easier for the analysis. The transformation is called conformal transformation, of which we change the scaling of the metric while keep the inner constructure unaltered. It has been proved that for pure $f(R)$ modified gravity theories or single scalar field nonminimally coupled to gravity, the curvature perturbations defined in the two frames are equivalent, and one can directly use what he gets in Einstein frame as his final result. However, for the models we are going to discuss about, as has been addressed in \cite{Gong:2011qe}, the curvature perturbations in two frames are no longer the same. As will be seen below, the difference is mainly due to the generation of isocurvature perturbations. Therefore, when the Einstein frame analysis is done, it is necessary to pull everything back to its Jordan frame to get the right results. In like manner, we also have to perform the constraint on observable in its Jordan frame with observational data.

Although some basic inflationary models has been checked against the observational data by the newest PLANCK Collaboration \cite{Ade:2013lta,Planck:2015xua}, many other models, including the ones we're talking about, were not. In this paper, we will also compare the power spectrum of curvature perturbation, in its Jordan frame, to the PLANCK data. Our numerical plot shows that it is very easy to make the theoretical results consistent with the observational data.

The present paper is organized as follows. Section II gives the background of our model. We transfer the model into Einstein frame and solve the equation of motion to get the inflationary solution in Einstein frame. In Section III, we study the perturbation theory of our model in detail in the Einstein frame and then pull back to the Jordan frame. One can see that the results of the two frames will be quite different. Section IV is devoted to the constraints on this model by use of PLANCK data. Finally, we conclude with a discussion in Section V. Note that, we will work with the reduced Planck mass, $M_{p} = 1/\sqrt{8\pi G}$, where $G$ is the gravitational constant, and adopt the mostly-plus metric sign convention $(-,+,+,+)$.

\section{The model}
We start with the action:
\be\label{action}
S=\int d^4\tilde{x} \sqrt{-\tilde{g}} \bigl[\frac{f(\tilde{R})}{2\kappa^2}+\mathcal{L}_s \bigr]~
\ee
with $\kappa^2=8\pi G$, where $f(\tilde{R})$ is an arbitrary function of the Ricci curvature $\tilde{R}$, and $\mathcal{L}_s=-\tilde{\nabla}_\mu\chi\tilde{\nabla}^\mu\chi/2-{\cal V}(\chi)$ is the lagrangian for the matter field $\chi$. For later convenience, we refer variables with tilde to those in Jordan frame while their correspondence without tilde are in Einstein frame. This is just a simple example of modified gravity accompanied with normal matter, but a big difference from pure modified gravity is that it contains more than one degree of freedom, where at large scales, the curvature perturbation is no longer a conserved variable but sourced by isocurvature perturbations, thus the system will not be in an adiabatic state. The equation of motion of the system (\ref{action}) can be written as:
\bea\label{freom1}
3\tilde{H}^2&=&\frac{1}{2F}(f+3F^{\prime\prime}+3\tilde{H}F^\prime)+\kappa^2\frac{\rho_s+3P_s}{2F}~,\\ \label{freom2}-2\tilde{H}^\prime&=&\frac{1}{F}(F^{\prime\prime}-\tilde{H}F^\prime)+\kappa^2\frac{\rho_s+P_s}{F}~,
\eea
where $\tilde{H}$ is the Hubble parameter, $F$ is defined as $F\equiv\partial f/\partial\tilde{R}$, and $\rho_s$ and $P_s$ are energy density and pressure of the scalar part, respectively. Prime denotes derivative with $\tilde{t}$, the cosmic time in Jordan frame. One can also define the ``effective" energy density and pressure for the whole system, which are
\bea
\tilde\rho&=&\frac{1}{2\kappa^2F}(f+3F^{\prime\prime}+3\tilde{H}F^\prime)+\frac{\rho_s+3P_s}{2F}~,\\ \tilde{P}&=&-\frac{1}{2\kappa^2F}(f+F^{\prime\prime}+5\tilde{H}F^\prime)+\frac{\rho_s-P_s}{2F}~,
\eea
satisfying the Friedmann equations $3\tilde{H}^2=\kappa^2\tilde\rho$ and $-2\tilde{H}^\prime=\kappa^2(\tilde{\rho}+\tilde{P})$.

As is well known, the system (\ref{action}) can be written into a Brans-Dicke form by simple field redefinition. Define $\varphi\equiv F$, $U(\varphi(\tilde{R}))=F\tilde{R}-f(\tilde{R})$, the action (\ref{action}) becomes:
\be
S_{BD}=\int d^4\tilde{x}\sqrt{-\tilde{g}} \biggl[\frac{\varphi\tilde{R}}{2\kappa^2}-U(\varphi)+\mathcal{L}_s \biggr]~,
\ee
which looks like two scalars, but still with one of them nonminimally coupled to gravity. The nonminimal coupling can be removed by further transformation, however, by the price of modifying the scaling of space-time, namely conformal transformation. To do this, we need to define a new metric, which we call the metric in Einstein frame. The metric in Einstein frame connects to the original metric (the metric in Jordan frame) as:
\be\label{conformal}
g_{\mu\nu}=\Omega^2\tilde{g}_{\mu\nu}~
\ee
where $\Omega\equiv\sqrt{\varphi}$. Therefore by manipulation we have:
\bea\label{einstein}
S_{E}&=&\int d^4x \sqrt{-g} \biggl[\frac{R}{2\kappa^2}-\frac{1}{2}\nabla_\mu\phi\nabla^\mu\phi\nonumber\\ &&-\frac{1}{2}e^{2b(\phi)}\nabla_\mu\chi\nabla^\mu\chi-V(\phi,\chi) \biggr]~,
\eea
where
\bea\label{phi}
\phi&=&-\sqrt{\frac{6}{\kappa^2}}\ln\Omega~,~b(\phi)=\sqrt{\frac{\kappa^2}{6}}\phi~,\nonumber\\
V(\phi,\chi)&=&\Omega^{-4}(U(\varphi)+{\cal V}(\chi))~.
\eea

From the above one can see, via conformal transformation one can transform our model (\ref{action}) into a minimal coupling two-field inflation model. The equation of motion for such two-field models contains two second-order differential equations, but no higher-order derivatives involving, thus is easier to be solved. That is a very important reason why people like to do the calculations in Einstein frame rather than directly in the original Jordan frame. For the system (\ref{einstein}), we would like to first analyze the background dynamics, and define some slow-roll parameters for later use. From (\ref{einstein}), The energy density and the pressure in Einstein frame are:
\bea
\rho&=&\frac{1}{2}\dot\phi^2+\frac{1}{2}e^{2b(\phi)}{\dot\chi}^2+V(\phi,\chi)~,\\
P&=&\frac{1}{2}\dot\phi^2+\frac{1}{2}e^{2b(\phi)}{\dot \chi}^2-V(\phi,\chi)~,
\eea
where dot denotes derivative w.r.t. cosmic time in Einstein frame $t$. By varying the action w.r.t. $\phi$ and $\chi$, we can obtain the equations of motion for both fields:
\bea\label{eom}
\ddot\phi+3H\dot\phi+V_\phi-b_\phi e^{2b(\phi)}{\dot\chi}^2&=&0~,\\
\ddot\chi+3H\dot\chi+2b_\phi\dot\chi\dot\phi+e^{-2b}V_\chi&=&0~.
\eea
Finally, the Friedmann equations read:
\be\label{friedmann}
H^2=\frac{\kappa^2}{3}\rho~,~\dot{H}=-\frac{\kappa^2}{2}(\rho+P)~.
\ee
Hereafter we take the unit such that $\kappa^2=1$. The various slow-roll parameters can be defined as:
\be\label{slowroll}
\epsilon_{\phi}=\frac{\dot\phi^2}{2H^2}~,~\epsilon_{\chi}=e^{2b}\frac{\dot\chi^2}{2H^2}~,~\epsilon=-\frac{\dot{H}}{H^2}~,
\ee
which should be much smaller than unity during inflation, and satisfy the relation: $\epsilon_\phi+\epsilon_\chi=\epsilon$. Under slow-roll approximation, the equation of motion (\ref{eom}) can be solved to give:
\be
\dot\phi\simeq-\frac{V_\phi}{3H}~,~\dot\chi\simeq-e^{-2b}\frac{V_\chi}{3H}~,~3H^2\simeq V(\phi,\chi)~.
\ee

\section{The perturbation of the model}
\subsection{Calculation of the perturbations in Einstein frame}
In this subsection, we first calculate the perturbations of our model in its Einstein frame, namely in form of a two-field inflation model (\ref{einstein}). The analysis of two-field inflation model has been well-developed and the detailed calculation can be found in e.g. \cite{Gong:2011qe,Gordon:2000hv,DiMarco:2002eb,Lalak:2007vi}, and here to be more concise we will only summarize their results which is needed for our later study. First of all, the perturbed metric can in general be formulated as:
\bea\label{perturbmetric}
ds^2&=&-(1+2\alpha)dt^2+2a(t)\partial_i\beta dtdx^i\nonumber\\
&&+a^2(t)[(1-2\psi)\delta_{ij}+2\partial_i\partial_j E]dx^idx^j~,
\eea
while the field can also be perturbed as
\be
\phi\rightarrow\phi_0(t)+\delta\phi(x)~,~\chi\rightarrow\chi_0(t)+\delta\chi(x)~.
\ee

It is convenient to define variables of perturbation that are invariant under gauge transformation. The often-used gauge invariant variables are:
\bea
\Phi&\equiv&\alpha-[a^2(\dot{E}-\beta/a)]^\cdot~,\\
\Psi&\equiv&\psi+a^2H(\dot{E}-\beta/a)~,\\
\cal R&\equiv&\psi-\frac{H}{\rho+P}\delta q~,\\
Q^i&\equiv&\delta\phi^i+\frac{\dot\phi^i}{H}\Phi~,~~~~(\delta\phi^i=\delta\phi,\delta\chi)~,
\eea
where $\Phi=\Psi$ is the Newtonian potential, $\cal R$ is the comoving curvature perturbation, and $Q^i$ is the gauge-invariant perturbation the $i$-th field. Moreover, $\delta q$ defined according to the relation $\partial_i\delta q=\delta T^0_i$ where $T^0_i$ is the $(0,i)$ component of the energy-momentum tensor of action (\ref{einstein}). In spatial-flat gauge which will be applied in this paper, one chooses $\psi=0$, and thus $Q^i$ identified with $\delta\phi^i$.

What we want to calculate is the comoving curvature perturbation $\cal R$, because it can be directly connected with the observables to test our model. However, what we can handle from the action (\ref{einstein}) is the field perturbations $Q^i$. Therefore, we need to connect between these two variables. One convenient technique is the so-called ``(instantaneous) Adiabatic-Entropy decomposition" \cite{Gordon:2000hv,Lalak:2007vi}, where one can decompose the field perturbations in the field-space $\{\dot\phi,\dot\chi\}$ into adiabatic and entropy directions, which traces along with/orthogonal to the field trajectory. Perturbations along the two directions are often called adiabatic and entropy perturbations of field respectively, which is proportional to the curvature perturbation $\cal R$ and the isocurvature perturbation, $\cal S$. Note that in single-field limit, there is only one field perturbation and the field-space is also one-dimensional, so the field perturbation can only goes along the field trajectory producing adiabatic perturbation, thus ${\cal S}\rightarrow 0$. In multi-field case however, isocurvature perturbations orthogonal to the trajectory can also exist, and as will be seen later, can act as a source of the adiabatic ones on large scales.

Following \cite{Gong:2011qe,Gordon:2000hv,DiMarco:2002eb,Lalak:2007vi}, we give the expression of adiabatic and entropy field perturbations ($Q^\sigma$ and $Q^s$) as:
\be\label{QsigmaQs}
Q^{\sigma}=\frac{\dot\phi}{\dot\sigma}\delta\phi+e^{2b}\frac{\dot\chi}{\dot\sigma}\delta\chi~, Q^s=-e^{b}\frac{\dot\chi}{\dot\sigma}\delta\phi+e^{b}\frac{\dot\phi}{\dot\sigma}\delta\chi~,
\ee
where $\dot\sigma=\sqrt{\dot\phi^2+e^{2b}\dot \chi^2}$ is the velocity of the background part of the adiabatic field $\sigma$. From the definition of $\cal R$ one can have
\be\label{curvature}
{\cal R}=\frac{H}{\dot\sigma}Q^\sigma~
\ee
in flat gauge. Similarly one can define the isocurvature perturbation as ${\cal S}=(H/\dot\sigma)Q^s$.

From action (\ref{einstein}) as well as Eq. (\ref{QsigmaQs}), one can finally obtain the equation of motion of $Q^\sigma$ and $Q^s$ after a long derivation \cite{Lalak:2007vi}:
\be\label{eomQsigma}
\ddot Q^\sigma+3H\dot Q^\sigma+(\frac{k^2}{a^2}+\mu_\sigma^2)Q^\sigma=-(\frac{2V_{,s}}{\dot\sigma} Q^s)^\cdot+\frac{2 V_{,\sigma}V_{,s}}{\dot\sigma^2} Q^s~,
\ee
where
\be
\mu_\sigma^2=-\frac{(\dot\sigma/H)^{\cdot\cdot}}{\dot\sigma/H}-(3H+\frac{(\dot\sigma/H)^{\cdot}}{\dot\sigma/H})\frac{(\dot\sigma/H)^{\cdot}}{\dot\sigma/H}~,
\ee
and
\be\label{eomQs}
\ddot Q^s+3H\dot Q^s+(\frac{k^2}{a^2}+\mu_s^2+\frac{4V_{,s}^2}{\dot\sigma^2})Q^s=\frac{2V_{,s}}{\dot H}\frac{k^2}{a^2}\Psi~,
\ee
where
\be\label{mus}
\mu_s^2=V_{,ss}+\frac{\dot\sigma^2}{2}R_F-\frac{V_{,s}^2}{\dot\sigma^2}~,
\ee
and the right hand side which is proportional to $k^2$ is negligible on large scales. Here $R_F$ is the scalar curvature in the {\it field space}, and in our model, we have $R_F=-2b_\phi^2$. Another useful relationship is between $V_{,s}$ and the slow-roll parameters, namely \cite{Lalak:2007vi}:
\be\label{Vs}
\frac{V_{,s}}{\dot\sigma}=H\eta_{\sigma s}-b_\phi\dot\sigma\sin^3\theta~.
\ee

By manipulating these equations, we can finally get a solution of $Q^\sigma$ and $Q^s$ as:
\be
Q^\sigma\simeq Q^s\sim\frac{\sqrt{-\tau}}{a(\tau)}H^{(1)}_{\frac{3}{2}}(-k\tau)~,
\ee
where we have taken the slow-roll limit and assume that the correlation between $Q^\sigma$ and $Q^s$ is small. $H^{(1)}_{3/2}$ is the first kind Hankel function of order $3/2$. In the $k\rightarrow\infty$ limit, from the above solution, one obtains the Bunch-Davies vacuum solution:
\be
Q^\sigma(\tau)\simeq Q^s(\tau)\simeq\frac{e^{-ik\tau}}{a(\tau)\sqrt{2k}}~
\ee
deep inside the horizon as usual, while in the $k\rightarrow 0$ limit, one gets
\be\label{largescale}
Q_\ast^\sigma\simeq Q_\ast^s\sim Hk^{-\frac{3}{2}}~
\ee
at the Hubble-exit time. Here $\tau\equiv\int a^{-1}(t)dt$ is the conformal time.

The inflationary observables are usually expressed in terms of power spectra and correlation functions. One could define the power spectrum for $Q^\sigma$ and $Q^s$ as:
\be
\langle Q^{m\dagger}Q^n\rangle=\frac{2\pi^2}{k^3}{\cal P}_m(k)\delta^{(3)}(\overrightarrow{k}-\overrightarrow{k}^\prime)\delta_{mn}~,~(m,n=\sigma,s)~
\ee
where $\delta_{mn}$ denotes that there are no correlations between $Q^\sigma$ and $Q^s$ until the Hubble-exit. This is actually reasonable, since inside the horizon, the correlations of the variables are determined by the communication relations of their production-annihilation operators, which should commute because they are independent degrees of freedom. Although they have coupling, it should be subdominant inside the horizon where $|k\tau|\gg 1$. Considering the relation between ${\cal R}$, ${\cal S}$ and $Q^\sigma$, $Q^s$ (Eq. (\ref{curvature}) and the sentences thereafter), one can have
\be
{\cal P}_{\cal R}=\frac{H^2}{\dot\sigma^2}{\cal P}_{\sigma}~,~{\cal P}_{\cal S}=\frac{H^2}{\dot\sigma^2}{\cal P}_{s}~,
\ee
and from the results (\ref{largescale}), one has:
\be\label{spectrum1}
{\cal P}_{{\cal R}\ast}\simeq{\cal P}_{{\cal S}\ast}=\frac{H_\ast^2}{8\pi^2\epsilon_\ast}~
\ee
at the Hubble-exit time. Here we denote the values of variables at Hubble-exit time by a star in the subscript. The spectral index of scalar spectrum at Hubble-exit time is then evaluated as
\be\label{index}
n_{{\cal R}\ast}\equiv\frac{d\ln{\cal P}_{{\cal R}\ast}}{d\ln k_\ast}=1-2\epsilon_\ast-\eta_\ast~.
\ee

One can also calculate the tensor perturbation of this model. In Einstein frame where the model behaves like two-field inflation model, the tensor power spectrum is the same as that of GR, which is
\be\label{tensor}
{\cal P}_{\cal T}=\frac{2H_\ast^2}{\pi^2}~.
\ee
One can also define the tensor-scalar ratio of the model, which is $r\equiv{\cal P}_{\cal T}/{\cal P}_{\cal R}$.

\subsection{Perturbations at large scales}
What we concerned more about is the values of perturbations that reenters the horizon. In single-field case, there is only curvature perturbation which is conserved at superhubble scales, so it is reasonable to take the Hubble-exit values of perturbation the same as those of the Hubble-reenter values, and thus the calculations in the above subsection is enough. However, in multi-field case, there is also isocurvature perturbation which will source the curvature one and the latter will evolve even after Hubble-exit, which we must take into account. We in this subsection give the formulation of perturbations at large scales as briefly as possible, while more detailed calculations can be found in the preceding works \cite{Lalak:2007vi,Langlois:2008wt}. From the equations of motion of the field perturbation (\ref{eomQsigma}) and (\ref{eomQs}), one can get the varying of the curvature and isocurvature perturbations at large scales as: \be\label{eqlargescale}
\dot{\cal R}\approx AH{\cal S}~,~\dot{\cal S}\approx BH{\cal S}~,
\ee
where
\be\label{AandB}
A=-\frac{2V_s}{H\dot\sigma}~,~B=-\frac{\eta}{2}-\frac{1}{3}(\frac{\mu_s^2}{H^2}+\frac{4V_s^2}{H^2\dot\sigma^2})~.
\ee
By integration over time one can get the expressions for ${\cal R}$ and ${\cal S}$ at late time, which is (in matrix form):
\be
\left(\begin{array}{c}
{\cal R}\\
{\cal S}\end{array}\right)=\left(\begin{array}{cc}
1 & T_{{\cal R}{\cal S}}\\
0 & T_{{\cal S}{\cal S}}\end{array}\right)\left(\begin{array}{c}
{\cal R}\\
{\cal S}\end{array}\right)_\ast~,
\ee
where
\bea\label{TrsandTss}
T_{{\cal R}{\cal S}}(t_\ast,t)&\equiv&\int^t_{t_\ast}A(t_1)T_{{\cal S}{\cal S}}(t_\ast,t_1)H(t_1)dt_1~,\nonumber\\
T_{{\cal S}{\cal S}}(t_\ast,t)&\equiv&\exp\left(\int^t_{t_\ast}B(t_2)(t_\ast,t_2)H(t_2)dt_2\right)~.
\eea

The power spectra at horizon-reentering therefore can be expressed as:
\be\label{spectrum2}
{\cal P}_{\cal R}=(1+T_{{\cal R}{\cal S}}^2){\cal P}_{{\cal S}\ast}~,~{\cal P}_{\cal S}=T_{{\cal S}{\cal S}}^2{\cal P}_{{\cal S}\ast}~,~{\cal C}_{{\cal R}{\cal S}}=T_{{\cal R}{\cal S}}T_{{\cal S}{\cal S}}{\cal P}_{{\cal S}\ast}~,
\ee
assuming that $\cal R$ and $\cal S$ are uncorrelated at Hubble-exit time. Define the rotation angle $\Theta$ such that $\sin\Theta=T_{{\cal R}{\cal S}}/\sqrt{1+T_{{\cal R}{\cal S}}^2}$, and applying the definition of spectral index in Eq. (\ref{index}), one can furtherly have:
\bea\label{index2}
n_{\cal R}&=&n_{{\cal R}\ast}-H_\ast^{-1}\sin(2\Theta)\frac{\partial T_{{\cal R}{\cal S}}}{\partial t_\ast}\nonumber\\ &=&1-2\epsilon_\ast-\eta_\ast-A_\ast\sin(2\Theta)-2B_\ast\sin^2\Theta~,
\eea
where the definition of $T_{{\cal R}{\cal S}}$ has been used. 

\subsection{Pulling back perturbations into Jordan frame}
In the previous sections we have calculated the perturbations of the model (\ref{action}), including the scalar and tensor perturbations, in its Einstein frame. This is what usually people do to analyse modified gravity theories, for it is much easier to deal with pure field theories in Einstein frame and one doesn't need to bother with the higher order curvature terms. For the case of pure modified gravity or single field nonminimal coupled with gravity, the curvature perturbation are invariant in both frames, i.e. $\tilde{\cal R}={\cal R}$ \cite{Makino:1991sg}, and one can directly use the Einstein frame results to compare with the observations. For our case, however, it is no longer the case. To see this, let's pull the results we get in Einstein frame back into the Jordan frame and see how different they are. From the conformal transformation (\ref{conformal}), we have:
\bea \label{relation}
\tilde{a}&=&\Omega_0^{-1} a~,~d\tilde{t}=\Omega_0^{-1} dt~,\nonumber\\
\tilde{H}&=&H\Omega_0-\dot\Omega_0=H\Omega_0-\Omega_0^\prime/\Omega_0~,
\eea
where $\Omega=\sqrt{\varphi}=\sqrt{F}$.

It is useful to define the Jordan-frame-based slow-varying parameter:
\bea \tilde\omega&\equiv&\frac{\Omega^\prime}{\tilde{H}\Omega}~,~\tilde\epsilon_\chi\equiv\frac{\chi^{\prime2}}{2\tilde{H}^2F}~,~\tilde\epsilon\equiv-\frac{\tilde{H}^\prime}{\tilde{H}^2}~, \nonumber\\
\tilde{z}&\equiv&\frac{\tilde{\omega}^\prime}{\tilde{H}\tilde{\omega}}~,\tilde\eta\equiv\frac{\tilde{\epsilon}^\prime}{\tilde{H}\tilde{\epsilon}}~,
\eea
where one could find from Eq. (\ref{freom2}) that first three of these parameters have the relation:
\be
\tilde\epsilon=-\tilde\omega+\tilde\epsilon_\chi+{\cal O}(\tilde\omega^2)~,
\ee
while we have $\epsilon_\phi\simeq 3\tilde{\omega}^2$ from Eqs. (\ref{phi}) and (\ref{slowroll}). Furthermore, one could get that the slow-roll parameters in the two frames are related as (up to leading order):
\be \epsilon=\tilde\epsilon+\tilde\omega\simeq\tilde\epsilon_\chi+3\tilde{\omega}^2~,~\eta=\frac{\tilde\eta\tilde\epsilon+\tilde{z}\tilde\omega}{\tilde\epsilon+\tilde\omega}~. \ee

Furthermore, from perturbed metric (\ref{perturbmetric}), one can have
\be
\tilde{\alpha}=\alpha-\frac{\delta\Omega}{\Omega_0}~,~\tilde{\psi}=\psi+\frac{\delta\Omega}{\Omega_0}~,
\ee
and the curvature perturbation in Jordan frame is:
\be
\tilde{\cal R}=\tilde{\psi}-\frac{\tilde{H}}{\tilde{\rho}+\tilde{P}}\tilde{\delta q}~.
\ee
From this expression, one can see that the curvature perturbation is not conformal invariant any more. To see this more clearly, one can calculate the difference of ${\cal R}$ of two frames, which is:
\bea\label{difference}
\tilde{\cal R}-{\cal R}&=&\tilde{\psi}-\frac{\tilde{H}}{\tilde{\rho}+\tilde{P}}\tilde{\delta q}-\psi+\frac{H}{\rho+P}\delta q~\nonumber\\ &=&\frac{\delta\Omega}{\Omega_0}+\frac{\tilde{H}}{2\tilde{H}^\prime}\tilde{\delta q}-\frac{H}{2\dot{H}}\delta q~,
\eea
where we have used the background equations. Note also that $\delta q$ is defined from the equation $\delta T^0_i=\partial_i\delta q$, so one has:
\bea
\tilde{\delta q}&=&-\frac{1}{F_0}(\chi^\prime\delta\chi+\delta F^\prime-F_0^\prime\tilde\alpha-\tilde{H}\delta F)~,\\
\delta q&=&-(\dot\phi\delta\phi+e^{2b}\dot\chi\delta\chi)~,
\eea
respectively.

Although the right hand side of Eq. (\ref{difference}) looks some complicated, it only contains terms that involves $\delta\phi$ and $\delta\chi$. Actually in such a two-field system (or $f(R)$+single field in Jordan frame), there are only two degrees of freedom and $\delta\phi$ and $\delta\chi$ can become a complete set which can present everything. Therefore, after some straightforward calculation and making use of the inverse transformation of Eq. (\ref{QsigmaQs}), we finally express $\tilde{\cal R}-{\cal R}$ in terms of $\delta\phi$ and $\delta\chi$ (or $\cal R$ and $\cal S$, because of their one-to-one correspondance) as: \be\label{differenceR}
\tilde{\cal R}-{\cal R}\simeq\Big[A\frac{\tilde\omega}{\tilde\epsilon}-\frac{(B-1)}{\tilde\epsilon}\sqrt{\frac{\tilde\epsilon+\tilde\omega}{3}-\tilde\omega^2}\Big]{\cal S}~,
\ee
where higher order terms are omitted and as it is perturbation on large scales, Eq. (\ref{eqlargescale}) is used.

From the above formula one can see that, the curvature perturbation ${\cal R}$ in the two frames differs by a quantity proportional to the isocurvature perturbation, ${\cal S}$. That means some parts of isocurvature perturbations has now been transferred into the adiabatic ones during frame transformation. So before and after transformation, we are actually talking about different ``adiabatic perturbations", which is in spite of the fact that mathematically the formulae in the two frames can be transformed to each other smoothly. Contrarily, in solo-degree-of-freedom (s.d.o.f.) system where the isocurvature perturbations do not appear, this difference will vanish and $\tilde{\cal R}$ and ${\cal R}$ will coincide, which gives the equivalence between curvature perturbations in two frames. Note that although we only show this point by taking a small example in this paper, in like manner, it can also be applicable for more complicated modified gravity models with m.d.o.f.. This conclusion is consistent with that in Ref. \cite{Gong:2011qe} obtained by using different methods.

According to (\ref{differenceR}), the power spectrum of $\tilde{\cal R}$ is:
\bea
\tilde{\cal P}_{\tilde{\cal R}}&\sim&|\tilde{\cal R}|^2~\nonumber\\
&\simeq&\Big|{\cal R}+\Big[A\frac{\tilde\omega}{\tilde\epsilon}-\frac{(B-1)}{\tilde\epsilon}\sqrt{\frac{\tilde\epsilon+\tilde\omega}{3}-\tilde\omega^2}\Big]{\cal S}\Big|^2~\nonumber\\
&\simeq&{\cal P}_{\cal R}+\Big[A\frac{\tilde\omega}{\tilde\epsilon}-\frac{(B-1)}{\tilde\epsilon}\sqrt{\frac{\tilde\epsilon+\tilde\omega}{3}-\tilde\omega^2}\Big]^2{\cal P}_{\cal S}\nonumber\\
&&+2\Big[A\frac{\tilde\omega}{\tilde\epsilon}-\frac{(B-1)}{\tilde\epsilon}\sqrt{\frac{\tilde\epsilon+\tilde\omega}{3}-\tilde\omega^2}\Big]{\cal C}_{{\cal R}{\cal S}}~,
\eea
where ${\cal P}_{\cal R}$, ${\cal P}_{\cal S}$ and ${\cal C}_{{\cal R}{\cal S}}$ can be related to their values at Hubble-crossing via relations (\ref{spectrum2}).

From the expressions of $A$ and $B$ in Eq. (\ref{AandB}) and making use of Eqs. (\ref{mus}) and (\ref{Vs}), it is straightforward to express $A$ and $B$ as:
\be\label{AandB}
A=-2\eta_{\sigma s}-b_\phi\sqrt{\epsilon}\sin^3\theta~,~B=-\frac{\eta}{2}-\frac{1}{3}\frac{V_{ss}}{H^2}+\frac{1}{3}b_\phi^2\epsilon-\frac{1}{4}A^2~,
\ee
and one can see that all the terms in $A$ and $B$ are slow-roll parameters. Therefore, although it is difficult to integrate (\ref{TrsandTss}) numerically, it is convenient to assume that the coefficients $A$ and $B$ in Eq. (\ref{eqlargescale}) are nearly constants, so one simply have:
\be
T_{{\cal R}{\cal S}}\simeq\frac{A}{B}(e^{BN_\ast}-1)~,~T_{{\cal S}{\cal S}}\simeq e^{BN_\ast}~,
\ee
where $N_\ast$ denotes the efolding number from Hubble-crossing time to the end of inflation. Taking this into account, one can straightforwardly get the final power spectrum of curvature perturbation in the original Jordan frame, namely $\tilde{\cal R}$, as:
\be\label{scalarspectrumJF}
\tilde{\cal P}_{\tilde{\cal R}}={\cal C}\frac{\tilde{H}_\ast^2}{8\pi^2F_\ast(\tilde\epsilon_\ast+\tilde\omega_\ast)}~.
\ee
where
\be
{\cal C}=1+\Big[\Big(\frac{A}{B}+\Big[A\frac{\tilde\omega_\ast}{\tilde\epsilon_\ast}-\frac{(B-1)}{\tilde\epsilon_\ast}\sqrt{\frac{\tilde\epsilon_\ast+\tilde\omega_\ast}{3}-\tilde\omega_\ast^2}\Big]\Big)e^{BN_\ast}-\frac{A}{B}\Big]^2~, \ee
and from Eq. (\ref{index2}) 
one can get the spectral index 
in Jordan frame as:
\bea\label{indexJF}
\tilde{n}_{\cal R}&=&1-2\tilde{\epsilon}_\ast-2\tilde{\omega}_\ast-\frac{\tilde{\eta}_\ast\tilde\epsilon_\ast+\tilde{z}_\ast\tilde\omega_\ast}{\tilde\epsilon_\ast+\tilde\omega_\ast}+\frac{{\cal C}^\prime}{\tilde{H}{\cal C}}~.
\eea
From Eqs. (\ref{scalarspectrumJF}) we can see that, when ${\cal C}=1$, $\tilde{\cal P}_{\tilde{\cal R}}$ will coincide with ${\cal P}_{{\cal R}}$, so ${\cal C}$ will behave as an estimator of the inequality between curvature perturbations in Einstein and Jordan frame in our model. From the expressions of $n_{\cal R}$ 
in 
Eq. (\ref{indexJF}), one could get nontrivial constraints on slow-roll parameters, which will be different from what is done in Einstein frame. The difference comes from two sources, one is due to the frame transformation, and the other is caused by the deviation of ${\cal C}$ from 1, namely the inequality of the perturbations in two frames.

How can we physically understand the possibility that the isocurvature modes do not decay to zero at the end of inflation? If that happens, it will indicate that there are still different components when inflation ends, namely the d.o.f.s in inflation may decay into different products. An example is that in m.d.o.f. systems, the dominant d.o.f. decays into radiation, while the subdominant ones decay into baryons, cold dark matter or neutrinos. As long as there has more than one component at the end of inflation, remnant isocurvature perturbations will appear, and according to Eq. (\ref{differenceR}), the frame difference of the adiabatic ones will show up. The newest PLANCK paper has carefully discussed these cases, and showed us the primordial isocurvature fractions $\beta_{iso}\equiv{\cal P}_{\cal S}/({\cal P}_{\cal R}+{\cal P}_{\cal S})$ in various cases by numerical study \cite{Ade:2013lta}. In the most general case, the $95\%$ upper limit of the primordial isocurvature fractions is $\beta_{iso}<0.6$, while in our models, for example when $T_{{\cal S}{\cal S}}$ takes the value $0.9$, one could get $\beta_{iso}\sim 0.4$, well within the allowed range. On the other hand however, the total vanish of the isocurvature modes means that the frame difference will disappear again at the end of inflation (although they have existed during inflation).

As a side remark, one may worry that the exponential form of the transfer function and $T_{{\cal S}{\cal S}}$ might enhance the isotropic fluctuations too much to conflict with the observations. However, if we scan its form more carefully, we will find that it is not the case. This is because that the factor $B$ will be quite small, namely only of order slow-roll parameters, while $N_\ast$ is nothing but the efolding number of the inflation, namely around the number $60$. As a special but explicit example, it has been numerically shown in Ref. \cite{Cai:2013caa} in which one of the two d.o.f.s act as curvaton, that the slow-roll parameters is around $10^{-3}$ to $10^{-6}$, making $T_{{\cal S}{\cal S}}\sim 0.9$. So no matter whether $B>0$ or $B<0$, the isocurvature perturbations will actually not expected to deviate much from the value at Hubble exit (although it does deviate). Moreover, since we can see that the most terms in $B$ are negative, there are large possibilities that $B$ is negative, meaning that the isocurvature modes are decaying. More exact identification of the sign of $B$ needs fitting at the whole parameter space, which is obviously beyond the scope of this paper.

It is also straightforward to get the tensor spectrum $\tilde{\cal P}_{\cal T}$ in Jordan frame of our model from Eq. (\ref{tensor}). After taking conformal tranformations (\ref{relation}) and keeping only leading order, the result will be
\be\label{tensorspectrumJF}
\tilde{\cal P}_{\cal T}=\frac{2\tilde{H}_\ast^2}{\pi^2F_\ast}~,
\ee
which coincides with ${\cal P}_{\cal T}$. This is because since tensor degrees of freedom of perturbations are decoupled from the scalar degrees of freedom, and in such a system there is only one tensor degree of freedom, so the tensor modes of the two frames are equivelant, and the tensor spectra in the two frames differ only by a conformal transformation. Eqs. (\ref{differenceR})-
(\ref{tensorspectrumJF}) are our main results in this context.
\section{Fitting with the PLANCK data}
In the previous section, we derived the perturbation generated by a ``modified gravity+single scalar" system, and after calculating in Einstein frame, we pull it back to its original Jordan frame, which can be used to fit the data. Actually, one can see that the results are controlled by a series of parameters, namely $\tilde{H}_\ast$, $\tilde{\epsilon}_\ast$, $\tilde{\eta}_\ast$, $\tilde{\omega}_\ast$, $\tilde{z}_\ast$, $F_\ast$ and ${\cal C}$, while ${\cal C}$ contains information after Hubble-exits, namely $A$, $B$ and $N_\ast$. Therefore, even we take explicit forms of $f(R)$ to break the degeneracy of some of the parameters, there are still large parameter space that can easily make our model consistent with the data. Here for illumination we just show two simple examples. In the first example $f(\tilde{R})$ is taken to be of the form \cite{DeFelice:2010aj}
\be
f(\tilde{R})=\xi\tilde{R}^n~,~F=n\xi\tilde{R}^{n-1}=n\xi[6\tilde{H}^2(2-\tilde\epsilon)]^{n-1}~,
\ee
and $\tilde\omega_\ast$ and $\tilde{z}_\ast$ can be simply expressed in terms of $\tilde{\epsilon}_\ast$, $\tilde{\eta}_\ast$ as
\be
\tilde\omega_\ast\simeq(1-n)\tilde{\epsilon}_\ast~,~\tilde{z}_\ast\simeq(1-n)\tilde{\eta}_\ast~,
\ee
and Eqs. (\ref{scalarspectrumJF}) and (\ref{indexJF})
becomes
\bea
\tilde{\cal P}_{\tilde{\cal R}}&=&{\cal C}\frac{\tilde{H}_\ast^{2(2-n)}}{8\times{12}^{n-1}\pi^2n(2-n)\xi\tilde{\epsilon}_\ast}~,\\
\tilde{n}_{\cal R}&=&1+2(n-2)\tilde{\epsilon}_\ast+(\frac{2}{n-2}+n)\tilde{\eta}_\ast+\frac{{\cal C}^\prime}{\tilde{H}{\cal C}}~
\eea
with
\bea
{\cal C}&=&1+\Big\{\frac{A}{B}[1+e^{BN_\ast}(nB-B-1)]\nonumber\\
&&+(B-1)e^{BN_\ast}\sqrt{\frac{2-n}{3\tilde\epsilon_\ast}-(n-1)^2}\Big\}^2~, \\
{\cal C}^\prime&=&2H_\ast e^{BN_\ast}\Big\{\frac{A}{B}[1+e^{BN_\ast}(nB-B-1)]\nonumber\\
&&+(B-1)e^{BN_\ast}\sqrt{\frac{2-n}{3\tilde\epsilon_\ast}-(n-1)^2}\Big\}\times\nonumber\\
&&\Big[A(nB-B-1)+(B-1)B\sqrt{\frac{2-n}{3\tilde\epsilon_\ast}-(n-1)^2}\nonumber\\ &&+\frac{(B-1)(n-2)}{6\sqrt{\frac{2-n}{3\tilde\epsilon_\ast}-(n-1)^2}}\frac{\tilde\eta_\ast}{\tilde\epsilon_\ast}\Big]~,
\eea
One can check that it returns to the results of standard inflation when $\xi=n=1$.

Another example is the well-known Starobinsky model \cite{Starobinsky:1980te}:
\be
f(\tilde{R})=\tilde{R}+\xi\tilde{R}^2~,~F=1+2\xi\tilde{R}=1+12\xi\tilde{H}^2(2-\tilde\epsilon)~,
\ee
and $\tilde\omega_\ast$ and $\tilde{z}_\ast$ can be simply expressed in terms of $\tilde{\epsilon}_\ast$, $\tilde{\eta}_\ast$ as
\be
\tilde\omega_\ast\simeq-\frac{24\xi\tilde{H}_\ast^2}{1+24\xi\tilde{H}_\ast^2}\tilde{\epsilon}_\ast~,~\tilde{z}_\ast=\frac{-2\tilde{\epsilon}_\ast}{1+24\xi\tilde{H}_\ast^2}+\tilde{\eta}_\ast~,
\ee
and Eqs. (\ref{scalarspectrumJF}) and (\ref{indexJF})
becomes
\bea
\tilde{\cal P}_{\tilde{\cal R}}&=&{\cal C}\frac{\tilde{H}_\ast^2}{8\pi^2\tilde\epsilon_\ast}~,\\
\tilde{n}_{\cal R}&=&1-2\tilde{\epsilon}_\ast-2\tilde{\eta}_\ast+\frac{{\cal C}^\prime}{\tilde{H}{\cal C}}~
\eea
with
\bea
{\cal C}&=&1+\Big\{\frac{A}{B}+e^{BN_\ast}\Big[A(1-\frac{1}{B}-\frac{1}{1+24\xi\tilde{H}_\ast^2})\nonumber\\ &&+\frac{(B-1)\sqrt{1+24\xi\tilde{H}_\ast^2(1-72\xi\tilde{H}_\ast^2\epsilon_\ast)}}{\sqrt{3\tilde\epsilon_\ast}(1+24\xi\tilde{H}_\ast^2)}\Big]\Big\}^2~,\\
{\cal C}^\prime&=&2H_\ast e^{BN_\ast}\Big\{\frac{A}{B}+e^{BN_\ast}\Big[A(1-\frac{1}{B}-\frac{1}{1+24\xi\tilde{H}_\ast^2})\nonumber\\ &&+\frac{(B-1)\sqrt{1+24\xi\tilde{H}_\ast^2(1-72\xi\tilde{H}_\ast^2\epsilon_\ast)}}{\sqrt{3\tilde\epsilon_\ast}(1+24\xi\tilde{H}_\ast^2)}\Big]\Big\}\times\nonumber\\
&&\Big[A(B-1-\frac{B}{1+24\xi\tilde{H}_\ast^2})-\frac{48A\xi\tilde{H}_\ast^2\tilde\epsilon_\ast^2}{(1+24\xi\tilde{H}_\ast^2)^2}\nonumber\\
&&+\frac{(B-1)B\sqrt{1+24\xi\tilde{H}_\ast^2(1-72\xi\tilde{H}_\ast^2\tilde\epsilon_\ast)}}{\sqrt{3\tilde\epsilon_\ast}(1+24\xi\tilde{H}_\ast^2)}+\nonumber\\ &&\frac{\sqrt{3}(B-1)}{6(1+24\xi\tilde{H}_\ast^2)^2\sqrt{\tilde\epsilon_\ast[1+24\xi\tilde{H}_\ast^2(1-72\xi\tilde{H}_\ast^2\tilde\epsilon_\ast)]}}\nonumber\\
&&\times\{48\xi\tilde{H}_\ast^2\tilde\epsilon_\ast[1+24\xi\tilde{H}_\ast^2(1+6\tilde\epsilon_\ast)]\nonumber\\
&&-(1+24\xi\tilde{H}_\ast^2)^2\tilde\eta_\ast\}\Big]~,
\eea
Interestingly, we find that in this case, although expressed using slow-roll parameters in Jordan frame, the expressions of $\tilde{n}_{\cal R}$ has the same form of that in Einstein frame, namely Eq. (\ref{index}), except for the ${\cal C}$ term. This holds for any value of $\xi$, and only depend on the quadratic scaling of the second term in $f(\tilde{R})$.

\begin{figure}[htbp]
\centering
\includegraphics[scale=0.4]{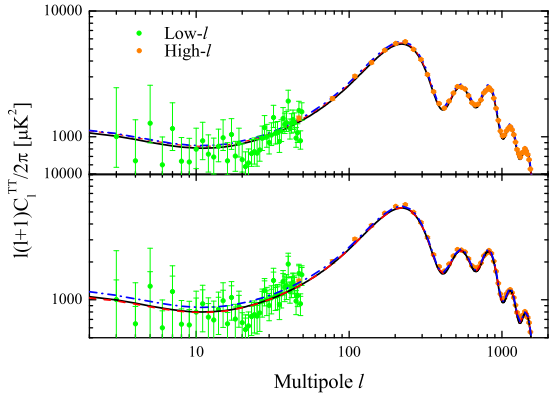}
\caption{The plot of TT spectrum from the examples in this section. The green and orange points and error bars are Planck data for low-$l$ and high-$l$, respectively. Upper panel: the $f(\tilde R)=\xi \tilde R^n$ case, where we choose $(\xi,n)$ as $(1,1.1)$ (black solid), $(10^4,1.8)$ (red dash), $(0.1,1.5)$ (blue dot dash) respectively. The power spectrum information $(A_s, n_s, r)$ are obtained as $(2.13\times10^{-9},0.9627,5.56\times10^{-3})$ (black solid), $(2.15\times10^{-9},0.9708,2.60\times10^{-2})$ (red dash), $(2.19\times10^{-9},0.9609,0.11)$ (blue dot dash). Lower panel: the $f(\tilde R)=\tilde R+\xi \tilde R^2$ case, where we choose $\xi$ as $1/6$ (black solid), $100$ (red dash), $1000$ (blue dot dash) respectively. The power spectrum information $(A_s, n_s, r)$ are obtained as $(2.17\times10^{-9},0.9629,2.73\times10^{-3})$ (black solid), $(2.21\times10^{-9},0.9703,1.55\times10^{-2})$ (red dash), $(2.27\times10^{-9},0.9645,1.28\times10^{-2})$ (blue dot dash).}\label{tt}
\end{figure}

Since the parameter space is still large for a global fitting, In Fig. \ref{tt} we take three cases of parameter choice for each example, and plot the TT power spectrum as well as compare them to the Planck data points. Although these three cases are randomly chosen and are different from each other, they are all consistent with the data, showing that it is quite easy to have our model fit the observations. Moreover, with proper choices, one of our cases tends to give a large tensor/scalar ratio $r$, which can contribute to TT power spectrum at large scales, and has possiblities to be justified by near future data. We present the parameter choices as well as the result such as $A_s$, $n_s$ and $r$ for each case in the caption of Fig. \ref{tt}. The newest PLANCK constraints on these quantities are $\ln(10^{10}A_s)\approx3.089\pm0.036$ and $n_s\approx0.9655\pm0.0062$ at $1\sigma$ confidence level, where PLANCK+LowP data are used \cite{Planck:2015xua}.

\section{conclusion}
In this paper we discussed about a well-motivated subclass of inflationary models, namely a scalar field in $f(R)$ modified gravity, and calculate the perturbations generated from this model. As a system of m.d.o.f., it is convenient to transform it into Einstein frame so that it becomes a minimal-coupling two-field system, the perturbations of which can be calculated in a standard way. However, contrary to the cases of pure $f(R)$ modified gravity or single field models nonminimally coupled to gravity in which the curvature perturbations in Jordan and Einstein frames are equivalent, in our model they are different. Therefore, one should ``pull-back" the Einstein-frame results via the conformal transformations in order to take the ``real" results of perturbations in the Jordan frame. In this paper, we calculated the power spectrum and spectral index of the curvature perturbations in Jordan frame, and showed their difference between those in Einstein frame. In the solo-degree-of-freedom system however, the quantities such as power spectrum and spectral indexes are given the same via computations in Jordan and Einstein frame, and one can directly use the results he gets in Einstein frame as his final results.

The main reason that causes the difference of the curvature perturbations in the two frames of our model is that, as a m.d.o.f. system, isocurvature perturbations will be generated, and the isocurvature perturbations in Einstein frame will contribute to the curvature ones in Jordan frame. In other words, although mathematically it has no problem to do such a transformation, physically we are concerning different quantities. So the difference of curvature perturbations between two frames are real, not only from conformal transformations. One can check that when the isocurvature fraction of the perturbations approaches zero, adiabatic perturbations in the two frames will again coincide. Our results are consistent with that in Ref. \cite{Gong:2011qe}, though the analysis are different. For more general systems of m.d.o.f., it is straightforward to prove that the conclusions are the same.

We also plot the TT spectrum according to the Jordan frame results we've got and compare it to the PLANCK data, with two simple examples. In the first one $f(R)$ takes the form of $R^n$, while the other is the famous Starobinsky model, $f(R)=R+\xi R^2$. From the plot one can see that, since in both examples the parameter space is quite large, it is very easy to have the theoretical plots of the spectrum consistent with the data points. We also explicitly gave the spectrum amplitude and spectral indexes of the two examples respectively, which is well within the constraints given by the PLANCK data.  Thus it also indicates that the model itself is interesting and deserves further investigations in the future.
\section{Acknowledgements}
We thank Yun-Song Piao, Antonio De Felice and Jonathan White for useful discussions. T.Q. is supported by the Open Innovation Fund of Key Laboratory of Quark and Lepton Physics (MOE), Central China Normal University (No.:QLPL2014P01), and J.X. is supported by the National Youth Thousand Talents Program.

\end{document}